\documentclass[conference]{IEEEtran}
\IEEEoverridecommandlockouts
\usepackage[T1]{fontenc}
\usepackage{tikz}
\usetikzlibrary{arrows}
\usepackage{todonotes}
\usepackage{cite}
\usepackage{amsmath,amssymb,amsfonts}
\usepackage{amsthm}
\usepackage{algorithmic}
\usepackage{textcomp}
\usepackage{xcolor}
\def\BibTeX{{\rm B\kern-.05em{\sc i\kern-.025em b}\kern-.08em
    T\kern-.1667em\lower.7ex\hbox{E}\kern-.125emX}}
\begin{document}



\newtheorem{definition}{Definition}[section]
\newtheorem{proposition}[definition]{Proposition}
\newtheorem{theorem}{Theorem}[section]
\newtheorem{axiom}{Axiom}[section]
\newtheorem{corollary}{Corollary}[theorem]
\newtheorem{lemma}[theorem]{Lemma}
\newtheorem{example}[definition]{Example}

\IEEEoverridecommandlockouts

\IEEEpubid{\begin{minipage}{\textwidth}\ \\[12pt]
979-8-3315-2931-4/24/\$31.00 \copyright 2024 IEEE
\end{minipage}}

\title{Building Intelligent Databases through Similarity: Interaction of Logical and Qualitative Reasoning
}

\author{\IEEEauthorblockN{
José-Luis Vilchis-Medina}
\IEEEauthorblockA{\textit{ENSTA Bretagne, Lab-STICC, UMR CNRS 6285} \\
Brest, France\\
jose.vilchis@ensta-bretagne.fr}
}

\maketitle

\begin{abstract}
In this article, we present a novel method for assessing the similarity of information within knowledge-bases using a logical point of view. This proposal introduces the concept of a similarity property space $\Xi_{\mathcal{P}}$ for each knowledge $\mathcal{K}$, offering a nuanced approach to understanding and quantifying similarity. By defining the similarity knowledge space $\Xi_{\mathcal{K}}$ through its properties and incorporating similarity source information, the framework reinforces the idea that similarity is deeply rooted in the characteristics of the knowledge being compared. Inclusion of super-categories within the similarity knowledge space $\Xi_{\mathcal{K}}$ allows for a hierarchical organization of knowledge, facilitating more sophisticated analysis and comparison. 
On the one hand, it provides a structured
framework for organizing and understanding similarity. The
existence of super-categories within this space further allows
for hierarchical organization of knowledge, which can be
particularly useful in complex domains. On the other hand, the
finite nature of these categories might be restrictive in certain
contexts, especially when dealing with evolving or highly
nuanced forms of knowledge. 
Future research and applications of this framework focus on addressing its potential limitations, particularly in
handling dynamic and highly specialized knowledge domains.
\end{abstract}


    \section{Introduction}
In today's information-rich world, the synergy between logic and qualitative reasoning has become increasingly vital. Logic forms the foundation of critical thinking, providing a structured approach to analyzing arguments and solving problems. Simultaneously, qualitative reasoning allows us to grasp the nuances of human experiences, interpreting context and emotions that quantitative data alone cannot capture. This combination proves crucial in decision-making processes~\cite{kuipers1994qualitative,brannen2017combining}. 
In the realm of knowledge-bases, a logical approach to data management has become increasingly crucial. One key aspect of this approach is the concept of searching data properties, which enables efficient retrieval and utilization of information within the knowledge-base.
The foundation of this logical approach lies in the recognition that data, when properly organized and structured, can serve as a powerful tool for knowledge discovery and decision-making~\cite{ceri1990logic,goldstone2012similarity}. 
Moreover, similarity logical approach can also help to manipulate the complexities of incomplete data. In a world where information is constantly evolving and new discoveries are made, knowledge-bases can often lag behind, leaving gaps and inconsistencies~\cite{libkin2014incomplete}. By embracing similarity as a guiding principle, we can bridge these gaps, extrapolating from the available data to fill the missing pieces and arrive at a more holistic understanding.
So, 
through the utilization of a similarity-based logical methodology, we can effectively harness the true potential of these repositories, thus transforming them into dynamic and adaptable tools capable of maintaining pace with the perpetually evolving landscape of information.

Additionally, the concept of similarity plays a critical role in how intelligent systems find patterns, make predictions, and enhance user experiences. Positive aspects of similarity can lead to more relevant recommendations and accurate classifications. However, the challenges of similarity come into play when considering superficial comparisons, which can inadvertently reinforce stereotypes or fail to recognize nuanced differences~\cite{noble1957psychology}. 
Similarity transcends mere resemblance, it is a pivotal factor that enables coordination, cooperation, synchronization, and recognition among diverse systems. From a computer science standpoint, the significance of similarity is manifested in its ability to address a wide range of challenges, particularly in domains such as robotics, autonomous vehicles, distributed databases, and networked systems~\cite{kuipers1994qualitative}. This relevance is underscored by its capacity to tackle critical issues that arise in these fields, including interoperability obstacles, the necessity for seamless collaboration, and the integration of heterogeneous technologies~\cite{hahn2001similarity,sheremet2007logic}.

As the digital age continues to reshape the way we interact with information, the importance of databases will only continue to grow. 
This work-in-progress aims to establish the basis for the study of information similarity from a logical point of view. 

    \subsection{Related works}
Intelligent Database Systems (IDS) exhibit several critical issues despite their sophistication. The inherent complexity of these systems often necessitates specialized knowledge and resources, resulting in steep learning curves and increased operational costs. Furthermore, their reliance on artificial intelligence algorithms renders them susceptible to biases and raises privacy concerns, particularly when handling sensitive information. The integration of these systems with machine learning models also heightens their vulnerability to cyber threats~\cite{doan2004ontology,kandola2002learning,devlin2018bert}.
The application of machine learning approaches to similarity measurement, while powerful, is heavily dependent on the quality and quantity of available data~\cite{minicozzi1976some}. If the data is poor or biased, the models may fail to generalize correctly. The selection of relevant features is a crucial aspect, and improper execution can lead to unsatisfactory results. Additionally, maintaining efficiency and speed in similarity computation for large datasets remains a significant challenge~\cite{severyn2015learning,huang2013learning,getoor2001learning}.

The utilization of Natural Language Processing (NLP) techniques for the assessment of semantic similarity is confronted with a set of inherent challenges~\cite{fellbaum1998wordnet,turney2010frequency,cer2018universal}. Many NLP methodologies rely heavily on the training corpus, and if this corpus is non-representative or biased, the resultant similarity measures may be misleading~\cite{miller1991contextual,raffel2020exploring}. The context-dependent similarity of words or phrases presents a particularly arduous obstacle, and models that fail to account for contextual factors may yield inaccurate outcomes.

Probabilistic approaches, including stochastic and Bayesian techniques, frequently depend on specific assumptions regarding the distribution of the data~\cite{cheng2010semantic,sullivan2015introduction,smith2013uncertainty,soize2017uncertainty}. When these assumptions are not met, the performance of the models can be compromised. The interpretation of the results obtained from these methods can be complex, thereby hindering the comprehension of similarity within specific contexts~\cite{tversky1977features, resnik1995using}. Furthermore, the similarity quantified through these methods may not be universally applicable across diverse domains~\cite{biazzo2002probabilistic,pearl1988probabilistic,mitra2005omen,hofmann1999probabilistic}.

Graph theory and ontology-based approaches, while offering improved interpretability, can become computationally intensive when dealing with large, complex datasets~\cite{gao2010survey,blondel2004measure,sowa2008conceptual}. Advanced methods in high-performance computing, such as entity alignment from knowledge graphs, require sophisticated computer systems with high memory capacity, limiting their accessibility~\cite{taieb2014ontol, shvaiko2011ontology, guarino2009overview, shu2007ontology}.
Given these limitations, a logic-based approach to knowledge similarity could offer a promising and innovative solution. Logic, with its foundation in formal reasoning and its ability to handle complex relationships, could address many of the shortcomings of current methods~\cite{sheremet2007logic,brachman1984tractability}.
A logic-based system for knowledge similarity could provide a more robust and transparent framework for similarity measurement. Unlike black-box machine learning models, logical systems can offer clear, interpretable reasoning paths~\cite{grosof2003description,brachman2004knowledge}. This transparency would be particularly valuable in domains where understanding the basis of similarity is crucial, such as in legal or medical applications.
In summary, a logic-based approach to knowledge similarity could serve as a viable alternative to the current computational constraints and lack of transparency in existing methods. This formal reasoning-based system could enhance the interpretability and accessibility of similarity measurement, particularly in critical domains where understanding the underlying reasoning is essential.

Furthermore, logic-based systems have the potential to overcome the data dependency challenges confronted by numerous contemporary methods. By relying on formal rules and relationships rather than extensive training datasets, these systems can provide more consistent and reliable similarity measures, even in domains where data is scarce or subject to bias.
The ability of logic to handle incomplete or uncertain information presents a significant advantage. Many real-world scenarios involve partial or uncertain knowledge, and a logic-based system can furnish more robust similarity measures in these situations, unlike probabilistic methods that often struggle with incomplete data.

In this article, we present the fundamental findings of a proposal aimed at evaluating the similarity or relatedness of distinct knowledge domains or concepts within a knowledge-base. The primary objective of this study is to introduce and discuss the core results of a logical approach designed to assess the similarity of knowledge.
The work is structured in two principal parts. The first section focuses on the formalization aspect, wherein a detailed explanation and a logical-mathematical model of the proposed methodology for assessing knowledge similarity are provided. The second part outlines the general conclusions drawn from the research findings and highlights the future work to be undertaken.

    \section{Formalisation}

    \subsection{Definitions}

    \begin{definition}\label{def:def3}
        All knowledge or (behavior) $\mathcal{K}$ involves properties
        $\mathcal{P}_{pro}$ described in First-Order logic (FOL).
        Formally:
        $\{ \bigwedge \mathcal{Q} \vdash \mathcal{P} \} \vDash \mathcal{P}_{pro} \
            \mid \ \mathcal{P}_{pro} \in \mathcal{K'} \ \mid \ \mathcal{K'} \subset
            \mathcal{K}$.
    \end{definition}

    \begin{example}\label{exa:example1}
        We can hypothetically consider two
        behaviors, $\{\mathcal{K}_{1} \cup \mathcal{K}_{2}\} \subset \mathcal{K}$, both of which have
        properties. Formally,
        $\{\mathcal{P}^{\mathcal{K}_{1}}_{1} \cup
        \mathcal{P}^{\mathcal{K}_{1}}_{2}\} \subset \mathcal{K}_{1}^{\mathcal{P}_{pro}}$,  
        $\{\mathcal{P}^{\mathcal{K}_{2}}_{1} \cup
        \mathcal{P}^{\mathcal{K}_{2}}_{2} \cup
        \mathcal{P}^{\mathcal{K}_{2}}_{3}\} \subset \mathcal{K}_{2}^{\mathcal{P}_{pro}}$.
        Moreover, properties are expressed in FOL: 
        $\{q_{1} \land q_{2} \vdash \mathcal{P}^{\mathcal{K}_{1}}_{1}\}$, 
        $\{q_{3} \land \neg q_{1} \vdash \mathcal{P}^{\mathcal{K}_{1}}_{2}\}$,
        $\{q_{4} \land q_{5} \vdash \mathcal{P}^{\mathcal{K}_{2}}_{1}\}$,
        $\{\neg q_{6} \land \neg q_{7} \vdash
        \mathcal{P}^{\mathcal{K}_{2}}_{2}\}$.
        $\{\neg q_{4} \land q_{6} \vdash \mathcal{P}^{\mathcal{K}_{2}}_{3}\}$.

    \end{example}

   \begin{definition}\label{def:def1}
        Any (known or unknown) knowledge $\mathcal{K}$ is expressed 
        in a declarative form.
    \end{definition}



    
    \begin{proposition}\label{thm:proposition1}
        For any given knowledge $\mathcal{K}$, there is 
        a \textit{similarity property
        space} $\Xi_{\mathcal{P}}$. 
        \begin{proof}
            From definitions \ref{def:def3} and \ref{def:def1}, 
            we know that
            any (known or unknown) knowledge $\mathcal{K}$ can be expressed as logical
            form: $\{\bigwedge
            \mathcal{Q} \vdash \mathcal{P}\}$. In addition, it is known that a
            logical form is composed of a body ($\mathcal{Q}$) and a head
            ($\mathcal{P}$), this form captures properties of an entity or
            situation. Thus, three groups for identifying similarities
            are defined. Namely, bodies and heads of proper properties
            ($\mathcal{P}_{pro}$) of a given knowledge are compared:
            \begin{equation}
                \begin{gathered}
                    \{\bigwedge \mathcal{Q} \vdash \mathcal{P}\} \vDash
                    \mathcal{P}_{pro} \ | \ \mathcal{P}_{pro} \subset
                    \mathcal{K}\\
                    \{\mathcal{Q}^{\mathcal{K}_{m}},\mathcal{P}^{\mathcal{K}_{m}}\}
                    \in
                    \mathcal{K}_{m}, \
                    \{\mathcal{Q}^{\mathcal{K}_{n}},\mathcal{P}^{\mathcal{K}_{n}}\}
                    \in
                    \mathcal{K}_{n} \ | \ \{\mathcal{K}_{m} \cup \mathcal{K}_{n}\}
                    \subset \mathcal{K}\\
                    \begin{split}
                        \Xi_{\mathcal{P}} & \supseteq 
                        \bigcup_{j=1}^{j \leq |\mathcal{P}^{\mathcal{K}_{m}}|}
                        \bigcup_{i=1}^{i \leq |\mathcal{P}^{\mathcal{K}_{n}}|}
                        \mathcal{P}_{j}^{\mathcal{K}_{m}} /
                        \mathcal{P}_{i}^{\mathcal{K}_{n}}
                    \end{split}
                    \\
                    \text{where }
                        \mathcal{P}_{j}^{\mathcal{K}_{m}} /
                        \mathcal{P}_{i}^{\mathcal{K}_{n}}
                        \text{ can have three cases:}
                        \\
                        \mathcal{K'}_{=}: \text{if
                        $
                        \{\forall \mathcal{Q}^{\mathcal{K}_{m}} \subseteq
                        \{\mathcal{Q}^{\mathcal{K}_{n}},
                        \mathcal{P}^{\mathcal{K}_{n}}\}\} \land 
                        \{\forall \mathcal{P}^{\mathcal{K}_{m}} \subseteq
                        \{\mathcal{Q}^{\mathcal{K}_{n}},
                        \mathcal{P}^{\mathcal{K}_{n}}\}\}
                        $};\\
                        \mathcal{K'}_{\approx}: \text{if
                        $
                        \{\exists \mathcal{Q}^{\mathcal{K}_{m}} \subset
                        \{\mathcal{Q}^{\mathcal{K}_{n}},\mathcal{P}^{\mathcal{K}_{n}}\}\}
                        \lor \{\exists \mathcal{P}^{\mathcal{K}_{m}} \subset
                        \{\mathcal{Q}^{\mathcal{K}_{n}},\mathcal{P}^{\mathcal{K}_{n}}\}\}
                        $};\\
                         \mathcal{K'}_{\neq}: \text{Otherwise.}
                    \\
                    \text{Then, }
                    \Xi_{\mathcal{P}} \supseteq
                        \{\mathcal{K}'_{=}\}^{|\mathcal{K}'_{=}|} \cup
                        \{\mathcal{K}'_{\approx}\}^{|\mathcal{K}'_{\approx}|}
                        \cup \{\mathcal{K}'_{\neq}\}^{|\mathcal{K}'_{\neq}|}
                \end{gathered}
            \end{equation}
        \end{proof}
    \end{proposition}

    \begin{corollary}\label{corollary:corollary2}
        Similarity knowledge space $\Xi_\mathcal{K}$ is defined by its properties.
        \begin{proof}
            From 
            Proposition~\ref{thm:proposition1}, similarity property space is
            formalized as follows:
            \begin{equation}
                \bigcup_{j=1}^{j \leq |\mathcal{P}^{\mathcal{K}_{m}}|}
                \bigcup_{i=1}^{i \leq |\mathcal{P}^{\mathcal{K}_{n}}|}
                \mathcal{P}_{j}^{\mathcal{K}_{m}} /
                \mathcal{P}_{i}^{\mathcal{K}_{n}}
                \subseteq
                \Xi_{\mathcal{P}}
            \end{equation}
            And from Definition~\ref{def:def3}, we have that knowledge involves
            properties: 
            $\mathcal{P}_{pro} \subseteq \mathcal{K}$.
            Keeping in mind that properties $\mathcal{P}_{pro}$
            are fundamental properties of a knowledge $\mathcal{K}$. Then
            properties
            $\mathcal{P}_{pro}$ of a given knowledge $\mathcal{K}_{}$ define the
            behavior of $\mathcal{K}_{}$. Thus for any property
            $\mathcal{P}_{j}$ of knowledge $\mathcal{K}_{m}$ must be the
            behavior of $\mathcal{K}_{m}$, in which it is the similarity
            knowledge space:
            \begin{equation}
                \bigcup_{i}^{i \leq |\mathcal{K}|}
                \bigcup_{\begin{tabular}{c}$j$\\$j \neq i$\end{tabular}}^{j \leq |\mathcal{K}|}
                \mathcal{K}_{j} /
                \mathcal{K}_{i}
                \subseteq
                \Xi_{\mathcal{K}}
            \end{equation}
\end{proof}
    \end{corollary}

    \begin{example}\label{exa:example2}
        Regarding knowledge $\{\mathcal{K}_{1} \cup \mathcal{K}_{2}\}
        \subseteq \mathcal{K}$ of Example~\ref{exa:example1} and
        using Proposition~\ref{thm:proposition1} in order to evaluate knowledge
        similarity, in other words, how similar is $\mathcal{K}_{1}$ to
        $\mathcal{K}_{2}$, so $\mathcal{K}_{1}/\mathcal{K}_{2}$. We know that
        knowledges are defined by properties, 
        $\{\mathcal{P}^{\mathcal{K}_{1}}_{1} \cup
        \mathcal{P}^{\mathcal{K}_{1}}_{2}\} \subseteq \mathcal{K}_{1}$,  
        $\{\mathcal{P}^{\mathcal{K}_{2}}_{1},
        \mathcal{P}^{\mathcal{K}_{2}}_{2},
        \mathcal{P}^{\mathcal{K}_{2}}_{3}\} \subseteq \mathcal{K}_{2}$. Thus, we can apply the function
        $\Xi_{\mathcal{P}}$ in order to genera the similarity properties
        space:
        \begin{equation}
            \bigcup_{j=1}^{j \leq 2} \bigcup_{i=1}^{i \leq
            3} \mathcal{P}_{j}^{\mathcal{K}_{1}} /
            \mathcal{P}_{i}^{\mathcal{K}_{2}} \subseteq
            \begin{Bmatrix}
                \mathcal{P}_{1}^{\mathcal{K}_{1}} /
                \mathcal{P}_{1}^{\mathcal{K}_{2}} & 
                \mathcal{P}_{2}^{\mathcal{K}_{1}} /
                \mathcal{P}_{1}^{\mathcal{K}_{2}} 
                \\
                \mathcal{P}_{1}^{\mathcal{K}_{1}} /
                \mathcal{P}_{2}^{\mathcal{K}_{2}} & 
                \mathcal{P}_{2}^{\mathcal{K}_{1}} /
                \mathcal{P}_{2}^{\mathcal{K}_{2}}
                \\
                \mathcal{P}_{1}^{\mathcal{K}_{1}} /
                \mathcal{P}_{3}^{\mathcal{K}_{2}} &
                \mathcal{P}_{2}^{\mathcal{K}_{1}} /
                \mathcal{P}_{3}^{\mathcal{K}_{2}}
            \end{Bmatrix}
            \subseteq \Xi_{\mathcal{P}}
        \end{equation}
        Arbitrarily we are going to consider that there are certain properties
        (both their bodies and their heads) which are the same, similar and
        different, according to formalization of the
        Proposition~\ref{thm:proposition1} for these cases.
        \begin{equation}
            \bigcup_{j=1}^{j \leq 2} \bigcup_{i=1}^{i \leq
            3} \mathcal{P}_{j}^{\mathcal{K}_{1}} /
            \mathcal{P}_{i}^{\mathcal{K}_{2}} \subseteq 
            \begin{Bmatrix}
                \mathcal{K}'_{\neq} & \mathcal{K}'_{\approx}
                \\
                \mathcal{K}'_{\approx} & \mathcal{K}'_{\approx}
                \\
                \mathcal{K}'_{=} & \mathcal{K}'_{\neq}
            \end{Bmatrix}
            \subseteq \Xi_{\mathcal{P}}
        \end{equation}
        And finally, we can describe the similarity property space
        $\Xi_{\mathcal{P}}$ in terms of cardinalities: 
        \begin{equation}
            \{ \{\mathcal{K}'_{=}\}^{1} \cup
            \{\mathcal{K}'_{\approx}\}^{3} \cup \{\mathcal{K}'_{\neq}\}^{2} \}
            \subseteq \Xi_{\mathcal{P}}
        \end{equation}
        Hence, knowledge $\mathcal{K}_{1}$ contains at least one equal property, three similar
        properties and 
        two different properties in relation to
        knowledge $\mathcal{K}_{2}$.
    \end{example}

    \begin{lemma}\label{lemma:lemma1}
        Similarity property space 
        $\Xi_{\mathcal{P}}$ contains similarity
        source information $\Xi_{\mathcal{K}}^{\dagger}$.
        \begin{proof}
            Given two or more knowledges,
            $\{\mathcal{K}_{1} \cup \mathcal{K}_{2} \cup
            \mathcal{K}_{3} \cup \ldots\} \subseteq \mathcal{K}$,
            with a constant (or non-constant) cardinality of properties of each, thus, a
            Cartesian product is performed in order to search similarities,
            $\mathcal{K}\times\mathcal{K}$. Consequently, we have a square size
            set, $| \mathcal{K} | \times | \mathcal{K} |$. 
            In order to discriminate redundancies, 
            we consider half of set: $2 \cdot
            \Xi_{\mathcal{K}}^{\dagger} \subseteq \Xi_{\mathcal{K}}$. Lastly, we can arbitrarily select the
            lower triangular space set without considering diagonal information 
            of the space. So, similarity source information
            $\Xi_{\mathcal{K}}^{\dagger}$ is located in the lower triangular space set of
            $\Xi_{\mathcal{K}}$.
        \end{proof}
    \end{lemma}

    \begin{example}\label{exa:example3}
        Let's consider 3 different kinds of knowledge
        $\{\mathcal{K}_1 \cup \mathcal{K}_2 \cup \mathcal{K}_3\} \subseteq \mathcal{K}$, and also
        that whole set of knowledge has equal
        cardinality of
        properties: $|\mathcal{K}_{1}^{\mathcal{P}_{pro}}| \subseteq
        |\mathcal{K}_{2}^{\mathcal{P}_{pro}}| \subseteq
        |\mathcal{K}_{3}^{\mathcal{P}_{pro}}|$.
        From Corollary~\ref{corollary:corollary2}, we have: 
        \begin{equation}
            \begin{gathered}
                \begin{split}
                \bigcup_{i}^{i \leq |\mathcal{K}|}
                \bigcup_{\begin{tabular}{c}$j$\\$j \neq i$\end{tabular}}^{j \leq |\mathcal{K}|}
                \mathcal{K}_{j} /
                \mathcal{K}_{i}
                \equiv
                \bigcup_{j=1}^{j \leq |\mathcal{P}^{\mathcal{K}_{m}}|}
                \bigcup_{i=1}^{i \leq |\mathcal{P}^{\mathcal{K}_{n}}|}
                \mathcal{P}_{j}^{\mathcal{K}_{m}} /
                \mathcal{P}_{i}^{\mathcal{K}_{n}}
                    \subseteq \Xi_{\mathcal{K}}
                \end{split}
            \end{gathered}
        \end{equation}
        And from Lemma~\ref{lemma:lemma1}, we know that similarity knowledge
        space $\Xi_{\mathcal{K}}$ has redundancies that can be expressed as
        follows: $2 \cdot \Xi_{\mathcal{K}}^{\dagger} \subseteq  \Xi_{\mathcal{K}}$.
        In order to select (lower or upper part) similarity source information 
        part $\Xi_{\mathcal{K}}$ we must not consider the diagonal of the space.
        \begin{equation}
            2 \cdot \Xi_{\mathcal{K}}^{\dagger} \subseteq
            |\mathcal{K} \times \mathcal{K}| - |\text{diag}(\mathcal{K})|
            \subseteq \Xi_{\mathcal{K}}
        \end{equation}
    \end{example}

    \begin{corollary}\label{corollary:corollary1}
        Similarity knowledge space $\Xi_{\mathcal{K}}$ is bounded by a finite number of categories.
        \begin{proof}
            We know that similarity knowledge space $\Xi_{\mathcal{K}}$ generates
            three different sets: 
            $\{\mathcal{K}'_{=}\}, \{\mathcal{K}'_{\approx}\},
            \{\mathcal{K}'_{\neq}\}$. Consequently, it will be eight different
            possible sets as result. Because of all the possible combinations
            from three sets, initializing with three empty sets up to three
            sets with information.
        \end{proof}
    \end{corollary}
    
    \begin{lemma}\label{lemma:lemma2}
        For any knowledge $\mathcal{K}$, similarity identification occurs when 
        knowledge similarity space
        $\Xi_{\mathcal{K}}$ is strictly non-empty: 
        \begin{proof}
            From Corollary~\ref{corollary:corollary1}, we know that there are eight
            possible categories. Then, non-empty set can only be obtained when all categories are non-zero cardinality:
            $\{\mathcal{K}'_{=}\} \cup \{\mathcal{K}'_{\approx}\} \cup 
            \{\mathcal{K}'_{\neq}\} \not \in \{\emptyset\}$.
            Consequently, categories out of this criterion can be considered for similarity identification.
        \end{proof}
    \end{lemma}

    \begin{lemma}\label{lemma:lemma3}
        There are super-categories in the knowledge similarity space
        $\Xi_{\mathcal{K}}$.
        \begin{proof}
            From Corollary~\ref{corollary:corollary1}, we know there are eight possible
            categories of similarities, strictly speaking we can actually consider only 7 by
            Lemma~\ref{lemma:lemma2}, where we discard null cardinality for the
            three categories case. 
            Consequently, we obtain three different super-categories: 
            These super-categories are the consequences of regrouping
            $\{\mathcal{K'}_{=} \cup
            \mathcal{K'}_{\approx} \cup \mathcal{K'}_{\neq}\}$ sets,
            which may be described as follows:
            \begin{itemize}
                \item \textbf{Case 1:} It will happen when two sets of
                    $\{\mathcal{K'}_{=} \cup
                    \mathcal{K'}_{\approx} \cup \mathcal{K'}_{\neq}\}$ are empty,
                    thus only 
                    $\{\mathcal{K'}_{=}\}$ or
                    $\{\mathcal{K'}_{\approx}\}$ or $\{\mathcal{K'}_{\neq}\}$
                    will remain. 
                \item \textbf{Case 2:} It will happen when one set of
                    $\{\mathcal{K'}_{=} \cup
                    \mathcal{K'}_{\approx} \cup \mathcal{K'}_{\neq}\}$ is empty, 
                    then three pairs of sets should be present:
                    $\{\mathcal{K'}_{=} \cup
                    \mathcal{K'}_{\approx}\}$ or $\{\mathcal{K'}_{=} \cup
                    \mathcal{K'}_{\neq}\}$ or $\{\mathcal{K'}_{\approx} \cup
                    \mathcal{K'}_{\neq}\}$.
                \item \textbf{Case 3:} It will occur when no empty set is
                    involved, resulting in a single set:
                    $\{\mathcal{K'}_{=} \cup
                    \mathcal{K'}_{\approx} \cup \mathcal{K'}_{\neq}\}$.
            \end{itemize}
        \end{proof}
    \end{lemma}

\subsection{Discussion}
The fundamental assertion that all knowledge involves properties describable in FOL is a powerful starting point. FOL's expressive power allows for the representation of complex relationships and structures, making it a suitable language for capturing the nuances of various forms of knowledge. This approach provides a standardized method for knowledge representation, potentially facilitating easier comparison and analysis across different domains.
The concept of a similarity property space $\Xi_{\mathcal{P}}$ for any given knowledge $\mathcal{K}$ is particularly interesting. It suggests that similarity is not an absolute measure but is relative to the specific properties relevant to the knowledge in question. This contextual approach to similarity aligns well with human intuition, what we consider similar in one context may not be in another.

The bounded nature of the similarity knowledge space $\Xi_{\mathcal{K}}$ by a finite number of categories is both a strength and a potential limitation. On one hand, it provides a structured framework for organizing and understanding similarity. The existence of super-categories within this space further allows for hierarchical organization of knowledge, which can be particularly useful in complex domains. On the other hand, the finite nature of these categories might be restrictive in certain contexts, especially when dealing with evolving or highly nuanced forms of knowledge.
The condition that similarity identification occurs only when the knowledge similarity space is strictly non-empty is a crucial point. It implies that for any meaningful comparison or similarity assessment to take place, there must be some shared property space between the knowledge entities being compared. This condition helps in avoiding spurious or meaningless similarity assessments.

However, this framework raises several questions and potential challenges:
\begin{itemize}
    \item Subjectivity in property selection: The selection of properties that define the similarity space could be subjective. Different observers might choose different properties, leading to varying similarity assessments.
    \item Dynamic nature of knowledge: Knowledge often evolves. How does this framework account for the dynamic nature of knowledge and the potential need for evolving similarity spaces?
    \item Handling uncertainty: FOL traditionally deals with definite truths. How does this framework handle uncertain or probabilistic knowledge?
\end{itemize}

    \section{Conclusion}
    This paper presented a new proposal for studying knowledge similarity using 
    a logical point of view. 
The introduction of a similarity property space $\Xi_{\mathcal{P}}$ for each knowledge entity $\mathcal{K}$ represents a nuanced approach to understanding similarity. By recognizing that similarity is context-dependent and relative to specific properties, this framework aligns closely with human intuition about how we perceive and compare concepts. The concept that the similarity knowledge space $\Xi_{\mathcal{K}}$ is defined by its properties and contains similarity source information further reinforces the idea that similarity is not an absolute measure but is deeply rooted in the characteristics of the knowledge being compared.

The bounded nature of the similarity knowledge space $\Xi_{\mathcal{K}}$ by a finite number of categories presents both opportunities and challenges. On one hand, it provides a structured framework for organizing and understanding similarity, which can be especially beneficial in complex domains. The inclusion of super-categories within this space allows for a hierarchical organization of knowledge, facilitating more sophisticated analysis and comparison. This hierarchical structure mirrors many natural and artificial classification systems, potentially making it more intuitive and applicable across various fields.


The condition that similarity identification occurs only when the knowledge similarity space $\Xi_{\mathcal{K}}$ is strictly non-empty is a critical aspect of this framework. This requirement ensures that meaningful comparisons are based on shared properties, preventing spurious or irrelevant similarity assessments. By enforcing this condition, the framework maintains the integrity of similarity comparisons, ensuring that they are grounded in substantive shared characteristics rather than superficial or coincidental similarities.

The existence of super-categories in the knowledge similarity space $\Xi_{\mathcal{K}}$ adds another layer of sophistication to this framework. It allows for a more nuanced understanding of how different knowledge entities relate to each other on multiple levels, potentially revealing higher-order similarities that might not be apparent at more granular levels of comparison.

    \bibliographystyle{splncs04}

    \bibliography{main}

\end{document}